\documentclass{llncs}

\usepackage{enumerate} 
\usepackage{enumitem} 
\usepackage[utf8]{inputenc} 
\usepackage{graphicx} 

\begin{document}

\title{The Emergence of Cooperation in Public Goods Games on Randomly Growing Dynamic Networks}



\author{Steve Miller\inst{1} \and Joshua Knowles\inst{2}}

\institute{School of Computer Science, University of Manchester, Manchester, UK
\and School of Computer Science, University of Birmingham, Birmingham, UK
\email{stevemiller.gm@gmail.com}}

\maketitle

\begin{abstract}

According to evolutionary game theory, cooperation in public goods games is eliminated by free-riders, yet in nature, cooperation is ubiquitous. Artificial models resolve this contradiction via the mechanism of network reciprocity. However, existing research only addresses pre-existing networks and does not specifically consider their origins. Further, much work has focused on scale-free networks and so pre-supposes attachment mechanisms which may not exist in nature. We present a coevolutionary model of public goods games in networks, growing by random attachment, from small founding populations of simple agents. The model demonstrates the emergence of cooperation in moderately heterogeneous networks, regardless of original founders’ behaviour, and absent higher cognitive abilities such as recognition or memory. It may thus illustrate a more general mechanism for the evolution of cooperation, from early origins, in minimally cognitive organisms. It is the first example of a model explaining cooperation in public goods games on growing networks.


\end{abstract}

\section{Introduction}

The prisoner's dilemma has become a standard metaphor to represent cooperation in evolutionary game theory, however it only describes interactions between \emph{pairs} of individuals.  In nature, interactions are not necessarily constrained in this way and a broader representation of cooperation is useful, particularly in the case of social, economic and biological networks~\cite{perc_evolutionary_2013}.  For such scenarios, the public goods game (PGG)  offers a suitable alternative for groups of more than two members.  Referred to variously, as the N-player prisoner's dilemma, the free-rider problem, or the tragedy of the commons~\cite{hardin_tragedy_1968}, the PGG represents a group-based dilemma where there exists a tension between benefits to an individual following one (selfish) course of action versus benefits to the entire community if the individual chooses an alternative action. 

The formulation of the PGG is  as follows:  Each member of a group has the opportunity to contribute a `cost' to a central `pot'.  They can choose to contribute, or not.  The amount invested in the pot is then increased by a multiplier.  The increased amount is divided amongst all members of the group, regardless of whether they contributed or not. Those contributing to the pot can be considered cooperators whilst those withholding, defectors (free-riders). As with the prisoner's dilemma, the choice which maximises payoff is to not contribute (to defect). Thus in the rational analysis, all individuals will choose to act selfishly, which will result in the worst case scenario for all: the minimisation of the public good. In nature however, the rational choice appears less appealing and communities are observed to cooperate, so as to preserve or maintain public goods. 

Attempts to explain this apparent contradiction between theory and observed behaviour consider the importance of factors such as volunteering, reputation, punishment or reward~\cite{hauert_volunteering_2002,brandt_punishment_2003,szolnoki_reward_2010}. Whilst it is easy to appreciate that such factors may affect the choices of, for example,  humans, higher primates or birds, it is harder to extend such approaches to explaining cooperation in more primitive forms of life~\cite{axelrod_evolution_1981} such as microorganisms cooperating to establish protective shelters, forage for nutrients or aid dispersal~\cite{crespi_evolution_2001}. In such cases, network reciprocity may offer an alternative explanation, requiring fewer assumptions.  

The effect of spatial structure on cooperation was first highlighted in~\cite{nowak_evolutionary_1992}. Whereas evolutionary game theory shows that cooperation cannot survive in evenly mixed populations, the presence of spatial relationships allows cooperators to cluster.  This clustering increases their individual fitnesses and thus prevents extinction by defectors. Further research developed these  findings to illustrate that heterogeneity of network structure promotes cooperation, in the case of \emph{pair-wise} interactions modelled using the prisoner's dilemma~\cite{santos_new_2006}. A similar approach modelling \emph{group-wise} interactions, using the PGG, was described in the work of~\cite{santos_social_2008}. Here, the mean field formulation of the PGG was spatially extended by mapping agents playing PGG  to nodes of a network. The results of this work illustrated the emergence of cooperation on scale-free networks, thus reinforcing previous findings regarding pair-wise (prisoner's dilemma) cooperation~\cite{santos_new_2006}.

Existing research has therefore established a consistent view of the positive role heterogeneous networks play in promoting cooperation, however the overwhelming majority of this work has focused on the pair-wise prisoner's dilemma and has primarily considered static networks. (A useful review of work focusing specifically on the PGG in networks may be be found in~\cite{perc_evolutionary_2013}). Of the limited body of research that exists for cooperation in dynamic networks, most has focused on networks at some form of equilibrium, using approaches which involve modification of pre-existing (fully formed) networks (see reviews in~\cite{szabo_evolutionary_2007,perc_coevolutionary_2010}). A very limited number of publications consider network growth~\cite{poncela_complex_2008,miller_population_2015,miller_minimal_2015}; all of the latter focusing on prisoner's dilemma.   

In this report we offer an initial attempt to fill this gap: We consider the growth of a population from its earliest origins and we ask how the social network affects and is affected by the \emph{group} behaviour of the individuals within it. Our aim is to establish a model based on group-wise cooperation which demonstrates the growth of networked populations of cooperative agents from original founder members. For such a model to be of value, it cannot be initially assumed that founder members are cooperators.  Further, for the model to be broadly applicable, the sort of cognitive abilities (memory, recognition, reasoning) that are required for reciprocity or retaliation cannot be assumed.  Finally for the model to be general, we make the simplest possible assumptions about the mechanism that new nodes use to attach 
to the existing network (i.e., we do not use preferential attachment).

\section{Background}

Here we discuss two models on which we have based our work and explain the rationale for the adaptations  made in incorporating these into a single model.  We provide this explanation in terms of the dynamic aspects of our model, divided into the two separate processes of attrition and growth.

\subsection{An Existing Network Representation of the Public Goods Game}
It has been demonstrated in~\cite{santos_social_2008} that heterogeneous network structure promotes cooperation in public goods games within static networks. The approach used represented a population in the form of multiple sub-groupings (neighbourhoods), each of which constitutes a PGG. More specifically, each node in the network initiates a single PGG and is also a participant in games initiated by its neighbours.  Hence each node takes part in $g = k+1$ games  where $k$ represents the degree of the node occupied by the agent. An agent $x$ with direct connections to neighbours $a$ and $b$ therefore has a degree of 2 and takes part in 3 PGGs: the one initiated by itself and the ones initiated by its neighbours $a$ and $b$. The total number of games in a population is therefore equal to $N$, the number of agents in the population. Within this work (ibid.), two variants of the PGG model were investigated, i) where each agent had a fixed cost per game (FCPG) and therefore their overall contribution was proportional to $g$, and ii) where each agent had a total fixed cost (fixed cost per individual, FCPI) and therefore their contribution was divided between all $g$ games. The game-playing populations are incorporated into evolutionary simulations by means of a strategy updating process representing natural selection.  Within this step, the strategies (behaviours) of fitter nodes probabilistically displace those of less fit neighbours.  

We aim to use the above approach as a basis from which to develop an extended dynamic model that simulates \emph{growth from founding members}. This naturalistic model is intended to explain the development of cooperation with respect to early origins of a population.

\subsection{An Existing Model of Cooperative Network Growth}
We take as our inspiration for developing a dynamic PGG model, the work of~\cite{poncela_complex_2008}, who notably connected the dynamic structure of a network to the behaviour of agents within the network.  In this approach, evolutionary processes, preferential attachment and agent behaviour were incorporated into a unified model of dynamic network-reciprocal cooperation (using the prisoner's dilemma), referred to as evolutionary preferential attachment (EPA).

\subsection{Proposed Attrition Mechanism}
In the EPA model, strategy updating still forms the primary evolutionary component, with selection acting on relative fitnesses resulting from \emph{agent-agent} interactions.  However, EPA also incorporates a secondary evolutionary mechanism into the \emph{growth} processes of the social \emph{network}.  Within our model, we shift this secondary evolutionary component over to \emph{shrinkage} of the network.  Specifically, this \emph{`global'} effect causes death of less fit \emph{individual agents}.  We consider this revision offers a model more analogous to the processes of selection in real world evolutionary situations.  Such a shift separates evolutionary effects from the attachment processes responsible for network growth. 

To implement such a culling mechanism, we impose a nominal maximum population size which is analogous to the concept of `carrying capacity', as used in population biology.  In this sense, the size of a population shrinks in response to extrinsic factors which are the result of environmental effects (such as predation, disease, food availability, many of which may be seasonal variations).

\subsection{Proposed Growth Mechanism}
The positive effect of scale-free degree distribution has featured significantly in research into the emergence of cooperation in networks~\cite{szabo_evolutionary_2007}.  However, we note that the fitness or degree-based mechanisms of preferential attachment, which are likely to be responsible for such structures, require underlying explanations for each occasion where they are found (for example, what specific process would enable a newcomer joining, or born into a population, to identify the fittest or most well-connected member in that situation).  Clearly preferential mechanisms exist (although disagreements have arisen over claims in this respect~\cite{clauset_power-law_2009,fox_keller_revisiting_2005}), however we suggest it is important that a general model for cooperation should be viable in the absence of mechanisms which require additional `case-by-case' explanations or assumptions (even if when present they may further enhance cooperation).  

To overcome these concerns, we implement the connection of new nodes to the existing network as an entirely random process.  Such a mechanism does not cause the development of a simple Poisson degree distribution as would be found in a random network: chronological random attachment (CRA) results in older nodes having more connections.  In the absence of other influences, the degree distribution in such a situation becomes exponential---giving a structure with heterogeneity somewhere between that of random and scale-free networks.

\subsection{Summary: A Model of Population Fluctuation in Social Networks}

The two processes described above, attrition of least fit nodes whenever a carrying capacity is reached, and growth of the network by random addition, continue until the simulation ends.  We thus have a fluctuation system which i) supports the growth of a network from founder members, and ii) overcomes the unrealistic situation that a `mature' network becomes fixed structurally. Further, as intended, this implementation gives us a minimal model which does not require assumption of higher cognitive abilities for its individual members and has no requirement regarding specific underlying mechanisms for the social network structure formation. This model, described in more detail below, is an extension of our earlier work~\cite{miller_population_2015}, which considered preferential attachment and the prisoner's dilemma game.

\section{Methods}

Our model describes agents located at the nodes of networks. Interactions occur between agents on nodes that have connecting edges.  Each node in the network has a `neighbourhood', defined by the neighbours its edges connect to. A PGG occurs for each neighbourhood and hence a network of $N$ nodes will result in $N$ PGGs. Agents can contribute to a PGG (cooperate) or not (defect). Each agent in the network has a behaviour encoded by a `strategy' variable representing either `cooperate' or `defect'. In a round robin fashion, each agent in turn initiates a PGG which involves their primary connected neighbours (their neighbourhood). Each agent in the population accumulates a fitness score which is the sum of its rewards from all the PGGs it participates in. 

Within the evolutionary simulation, this process is repeated over generations. Agents are assessed at each generation, on the basis of their fitness score: Fitter agents' strategies remain unchanged; less fit agents are more likely to have strategies displaced by those of fitter neighbours.  Fluctuation of the population occurs by repeated attrition and regrowth of the network. 

The general outline of the evolutionary process, for one generation, is as follows:

\begin{enumerate}[noitemsep]
	\item \emph{Play public goods games}: Each agent initiates a PGG involving its neighbours.  Each agent will accumulate a fitness score that is the sum of payoffs from all the individual PGGs that it participates in.
	\item \emph{Update strategies}: Selection occurs. Agents with low scores will have their strategies replaced, on a probabilistic basis, by comparison with the fitness scores of randomly selected neighbours.
	\item \emph{Remove nodes}: If the network has reached the nominal maximum size, it is pruned by a tournament selection process that removes less fit agents.
	\item \emph{Grow network}: A specified number of new nodes are added to the network, each connecting to $m$ randomly selected distinct existing nodes via $m$ edges. 

\end{enumerate}

\noindent In the following, we provide more detail on the specifics of each of the four steps:\\

\noindent \textbf{\emph{Play public goods games}}. Each node of the network, in turn, initiates a PGG.  Within a single PGG, all cooperator members of a neighbourhood contribute a cost $c$ to `the pot'.  The resulting collective investment $I$ is multiplied by $r$, and $rI$ is then divided equally amongst all members of the neighbourhood, regardless of strategy.

In the FCPG variant of the PGG, each agent has a fixed cost \emph{per game} and therefore their overall contribution, in one generation, is $c(k+1)$ with contribution $c$ to each game, and where $k$ is the number of neighbours (degree). The single game individual payoffs of an agent $x$ are given by the following equations, for scenarios where $x$ is a defector ($P_D$) and a cooperator ($P_C$) respectively:

\begin{equation}
P_D = crn_c/(k_x+1) \enspace,
\label{eqn:FCPG_defector_scoring}
\end{equation}

\begin{equation}
P_C = P_D-c \enspace,
\label{eqn:FCPG_cooperator_scoring}
\end{equation}

\noindent where $c$ is the cost contributed by each cooperator, $r$ is the reward multiplier, $n_c$ is the number of cooperators in the neighbourhood based around $x$, and $k_x$ is the degree of $x$.

In the FCPI variant, each \emph{individual} has a fixed cost $c$, i.e. their overall contribution is $c$ and hence their contribution to each game is $c/(k+1)$. The single game individual payoff for a node $y$ having strategy $s_y$ ($= 1$ if cooperator, $= 0$ if defector) present in the neighbourhood of $x$ is given by:

\begin{equation}
P_{y,x} = \frac{r}{k_x+1}\sum\limits_{i=0}^{k_x}\frac{c}{k_i+1}s_i - \frac{c}{k_y+1}s_y \enspace,
\label{eqn:FCPI__scoring}
\end{equation}

where $i$ is used to index each neighbour of $x$, and $s_i$ is the strategy of neighbour $i$ of $x$ having degree $k_i$.

\smallskip

\noindent \emph{\textbf{Update strategies}}. Each node $i$ selects a neighbour $j$ at random. If the fitness of node $i$, $f_i$ is greater or equal to the neighbour's fitness $f_j$, then $i$'s strategy is unchanged. If the fitness of node $i$, $f_i$ is less than the neighbour's fitness, $f_j$, then $i$'s strategy is replaced by a copy of the neighbour $j$'s strategy, according to a probability proportional to the difference between their fitness values. Thus poor scoring nodes have strategies displaced by those of more successful neighbours. 

Hence, at generation $t$, if $f_{i}(t)\geq f_{j}(t)$ then $i$'s strategy remains unchanged. If $f_{i}(t)< f_{j}(t)$ then $i$'s strategy is replaced with that of the neighbour $j$ with the following probability:

\begin{equation}
	\Pi_{U_i}(t) = \frac
	{f_j(t) - f_i(t)}
	{\max(k_i(t),k_j(t))} \enspace,
	\label{eqn:strategy_updating}
\end{equation}

\noindent where $k_{i}$ and $k_{j}$ are degrees of node $i$ and its neighbour $j$ respectively. The purpose of the denominator is to normalise the difference between the two nodes. The term $max(k_{i}(t),k_{j}(t))$ represents the largest achievable fitness difference between the two nodes given their respective degrees.

\smallskip

\noindent \emph{\textbf{Grow network}}. New nodes, with randomly allocated strategies, are added to achieve a total of 10 at each generation. Each new node uses $m=2$ edges to connect to existing nodes.  Duplicate edges and self-edges are not allowed. The probability $\Pi_{G_i}(t)$ that an existing node $i$ receives one of the $m$ new edges is given by: 

\begin{equation}
	\Pi_{G_i}(t) = \frac {1} {N(t)} \enspace,
	\label{eqn:RCA_node_addition}
\end{equation}

\noindent where  $N(t)$ is the number of nodes available to connect to at time $t$ in the existing population.   Given that in our model each new node extends $m = 2$ new edges, and multiple edges are not allowed, $N$ is therefore sampled \emph{without replacement}. Growth continues until a nominal maximum size (we used 1000 nodes) is achieved.

\smallskip

\noindent \emph{\textbf{Remove nodes (for fluctuation simulations)}}. On achieving or exceeding the nominal maximum size, the network is pruned by a percentage $X$. This is achieved by repeated tournament selection using a tournament size equivalent to $1\%$ of the population. Tournament members are selected randomly from the population. The tournament member having the least fitness is the `winner' and is added to a short list of nodes to be deleted. Tournament selection continues until the short list of $X\%$ nodes for deletion is fully populated. 

The nodes on the short list (and all of their edges) are removed from the network. Any nodes that become isolated from the network as a result of this process are also deleted. (Failure to do this would result in small numbers of single, disconnected, non-playing nodes, having static strategies and zero fitness values.) When there are multiple nodes of equivalent low fitness value, the selection is effectively random (on the basis that the members were originally picked from the population randomly). Where $X = 0$, no attrition occurs; in this case, on reaching maximum size, the network structure would become static.\\

\noindent\textbf{General simulation conditions}. Initial strategy types of founder nodes were specified in simulation setup (either 3 cooperators or 3 defectors).  Strategy types of subsequently added nodes were allocated independently, uniformly, at random (cooperators and defectors with equal probability). All networks had an overall average degree of approximately $k = 4$, giving an average neighbourhood size of $g = 5$. Simulations were run until 20,000 generations. The final `fraction of cooperators' values we use are means, averaged over the last 20 generations of each simulation, in order to compensate for variability that might occur from just using final generation values. Each simulation consisted of 25 replicates. We used shrinkage value of $X$ = 2.5\% for all fluctuation simulations. Simulation data is recorded after step 2 (\emph{Update strategies}).

\section{Results and Discussion}
We now present the results of research investigating our model's ability to support cooperation in a range of simulations. Initially we consider its implementation in the type of scenarios that have dominated research into cooperation in dynamic networks, namely fully formed or `pre-existing' networks.  We then apply the model, as per our original motivation, to consider networks grown from a small number of founding members.  In both types of investigation we have considered the two variants of PGG described in~\cite{santos_social_2008}: FCPG and FCPI.

\subsection{Simulations using Pre-existing Networks}

We first consider the impact of our model in pre-existing networks. We initially consider the effect of the PGG variant (FCPG vs. FCPI) and subsequently we discuss specifically how the fluctuation mechanism achieves different outcomes to those seen for static networks.

\subsubsection{Effect of PGG Variant in Simulations using Pre-existing Networks. }

It is established for the 2-player PGG (the prisoner's dilemma) that cooperation in pair-wise interactions on networks is promoted by the opportunity for cooperators to self-assort and form inter-connected groups (clusters)~\cite{nowak_evolutionary_1992}. The larger the clusters which form, the greater the levels of cooperation which occur~\cite{santos_new_2006}. This effect is therefore enhanced by increased network degree heterogeneity, since greater heterogeneity allows for increasingly larger connected groups of cooperators within the network. Such findings for the prisoner's dilemma generalise to the PGG, however in the PGG there is also an opposing `force' that limits cooperation, which we now explain. In the conventional FCPG representation of the PGG, an individual pays a cost for every single game they participate in.  Since each individual in a population can initiate a PGG among their local neighbourhood, higher connectivity (more neighbours) means that an individual will participate in more games and will thus pay a penalty for their increased connectivity~\cite{boyd_evolution_1988}. The classical result for the PGG in this case, is that the larger the neighbourhoods become, the less likely cooperation is. This finding makes intuitive sense, since the larger a PGG neighbourhood is, the closer it gets to representing a mean field scenario, where defection is the Nash equilibrium. 

\begin{figure}[!h]
	\begin{center}
		\includegraphics[width=12.2cm]{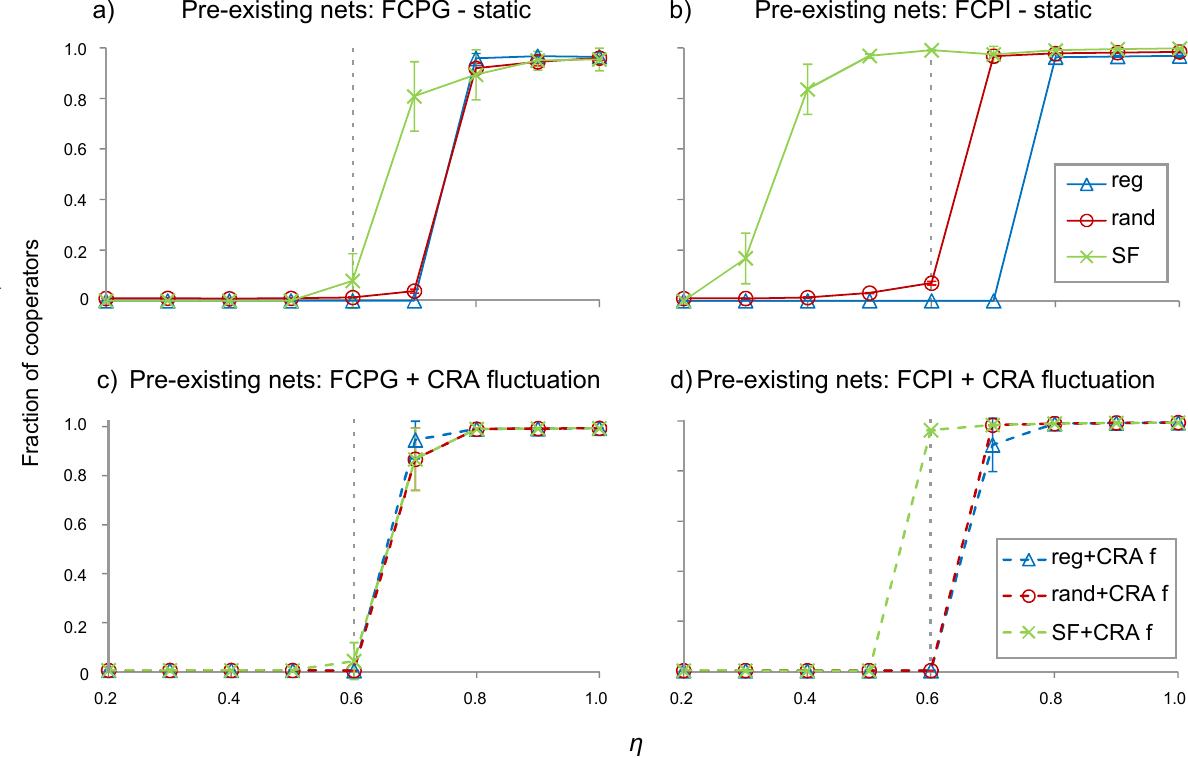}
		\caption{Plots comparing the effect of network type for simulations (25 replicates) on pre-existing networks of increasing heterogeneity (regular, random and scale-free respectively).   Final fraction of cooperators in population  is plotted against $\eta$, the PGG reward multiplier $r$, normalised with respect to average neighbourhood size ($g = 5$). Variability is indicated by error bars showing 95\% confidence intervals. Simulation details are as described in Methods section. The dashed line at $\eta = 0.6$ is a reference for the eye. }
		\label{fig:Effect_FCPG_v_FCPI}
	\end{center}
\end{figure}

In Fig. \ref{fig:Effect_FCPG_v_FCPI}a, (FCPG in static networks) we see that higher levels of cooperation are observed in static scale-free networks (green line with `x' markers) than in networks of low or no heterogeneity---random and regular respectively. These results are consistent with the view that heterogeneity promotes cooperation. In the case of FCPI PGG (see Fig. \ref{fig:Effect_FCPG_v_FCPI}b), the lack of any penalty on larger  neighbourhood size weakens the dilemma i.e. it reduces  the `temptation to defect' and therefore increases levels of cooperation.  By comparing corresponding lines for FCPG and FCPI PGG in Figs. \ref{fig:Effect_FCPG_v_FCPI}a and b, we can see how FCPI causes different horizontal shifts in cooperation profiles for networks of differing heterogeneity. We thus see that the impact of FCPI is nonexistent for regular networks (no visible shift, see blue lines with triangle markers) and becomes more relevant as increasing heterogeneity allows for increasing neighbourhood size (marked shift for scale-free networks, see green lines with `x' markers).

Figures \ref{fig:Effect_FCPG_v_FCPI}c and d illustrate the effect of PGG variant in fluctuating networks. Here we expect to see two general results: i) Given that the fluctuation mechanism drives all networks to the same final degree distribution, we would expect similar result profiles, within each variant of the PGG, regardless of starting network topology; ii) Further, we would expect these profiles, based on approximately exponentially distributed final networks, to lie somewhere between the two extremes of heterogeneity represented by scale-free (highly heterogeneous) and regular (non heterogeneous) networks, as observed in the static FCPI results (see Fig. \ref{fig:Effect_FCPG_v_FCPI}b). 

We find that the CRA-fluctuation profiles do indeed lie within the expected region of the graph, however we see an anomaly for the scale-free FCPI result (Fig. \ref{fig:Effect_FCPG_v_FCPI}d) which achieves higher levels of cooperation than regular and random networks. This result is unexpected because given an assumption that cooperation is only dependent on the final degree distribution, we would expect to see the same result profiles for all network types. We have compared final degree distributions for all network types and find no discernible difference. 

We propose that the explanation for the anomaly seen for initially scale-free networks lies in the differing challenges presented by the topology of the initial networks; specifically, the diameter of the network (rather than the degree distribution). We have measured the average shortest path length in our initial networks and find these to be approximately: 125 for regular networks, infinite (network disconnected) for random networks , and 4 for scale-free networks. Final networks have lengths of 6 in the case of FCPG and 7 in the case of FCPI. In order for cooperation to percolate through the network, sufficient reward ($\eta$) has to be present to drive assortativity by strategy, however assortativity will inevitably be impeded in those cases where the network is fragmented or does  not have the small path lengths that are a defining characteristic of small-world networks.  In such cases, cooperation cannot readily percolate, until the fluctuation mechanism has brought about sufficient changes to reduce the average path length and/or the number of network components. Whilst cooperation in scale-free networks is still dependent upon the value of $\eta$, such networks do not have to overcome the path length issues faced by random and regular networks. Thus, while all network topologies end up with the same final degree distributions, scale-free networks potentially start with a `small-world' advantage which may support the emergence of cooperation at lower values of $\eta$.

This proposed explanation raises the question of why a difference exists between scale-free network results for FCPG and FCPI (see Figs. \ref{fig:Effect_FCPG_v_FCPI}c and d).  In response to this,  our above explanation does indeed apply to fluctuating scale-free networks for both variants of the PGG, however in the case of FCPG, cooperation is limited by the additional constraint of neighbourhood size.

\subsubsection{Effect of Static vs. Fluctuating Networks.}

As reported in the work of~\cite{miller_minimal_2015}, a model based on fluctuating population size can promote cooperation in networks. This outcome arises from the greater opportunity for strategies to self-assort, given repeated perturbation of network structure, due to deletions of low fitness nodes.  

\begin{figure}[!h]
	\begin{center}
		\includegraphics[width=12.2cm]{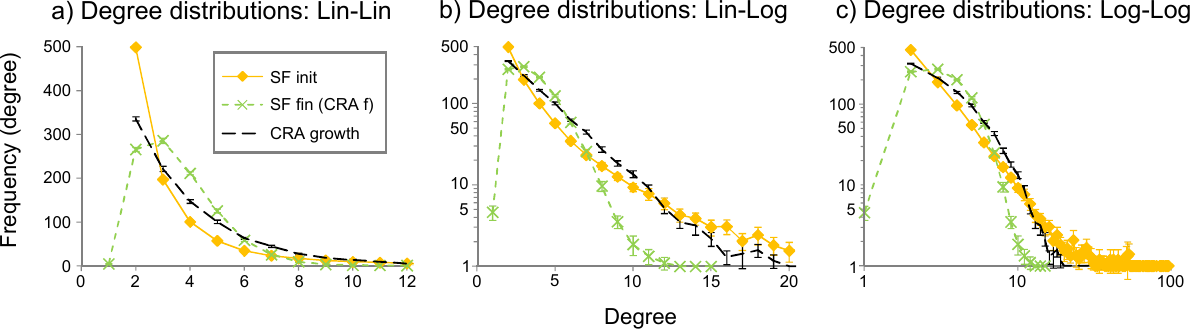}
		\caption{\textbf{a}) Graph illustrating initial and final degree distributions (25 replicates) for an initially scale-free network, of 1000 nodes, subjected to CRA-fluctuation. As a reference, the dashed black line represents the exponential distribution of a network `built' by CRA---without any node deletion.  To highlight characteristic distribution profiles, the same data is also presented on, \textbf{b}) lin-log (exponential appears as a straight line), and \textbf{c}) log-log (scale-free appears as straight line) scales.  For clarity, the x axis is truncated for plots \textbf{a} and \textbf{b}. Data is shown with error bars representing 95\% confidence intervals.  (The presence of small numbers of nodes of degree 1 is an artefact of our implementation of fluctuation: Whilst all new nodes initially have $m=2$ edges, the deletion process may leave limited numbers of nodes with a single edge.) }
		\label{fig:SF_before_n_after}
	\end{center}
\end{figure}

From comparing Figs. \ref{fig:Effect_FCPG_v_FCPI}a and c, we see how the incorporation of population fluctuation affects results for FCPG. Profiles for all network types are now superimposed. We observe similar findings in the case of FCPI (compare Figs. \ref{fig:Effect_FCPG_v_FCPI}b and d) except for in the case of scale-free networks---an anomaly which, as explained earlier, is believed to be due to the `beneficial' impact of short average path lengths found in initially scale-free networks. The \emph{general} consistency of cooperation profiles is because the CRA-fluctuation mechanism converts all networks to the same final degree distribution regardless of initial topology. Figure \ref{fig:SF_before_n_after} illustrates this conversion by showing initial and final degree distributions in simulations starting from a scale-free network. We note that whilst the final degree distribution due to CRA alone would be exponential, the additional effect of node deletion compresses the exponential curve, giving a degree heterogeneity lying between that of a Gaussian and an exponential distribution.  

We highlight an important point here. CRA-fluctuation converts \emph{all} networks, regardless of initial type to this compressed exponential distribution which has moderate heterogeneity.  In the case of scale-free networks, fluctuation therefore brings about a  \emph{decrease} in heterogeneity.  Hence for initially scale-free networks, we should expect to see reduced cooperation in the fluctuation model compared to the static one.  This is indeed true for FCPI results (green lines with `x' markers in Figs. \ref{fig:Effect_FCPG_v_FCPI}b and d) but not the case for FCPG (Figs. \ref{fig:Effect_FCPG_v_FCPI}a and c).  Fluctuation does not cause the expected reduction in FCPG PGG because in the static FCPG implementation, scale-free networks are \emph{already} constrained in their ability to cooperate i.e. they  cannot achieve their full potential in supporting cooperation due to the penalties FCPG imposes on large neighbourhoods.  

\begin{figure}[!h]
	\begin{center}
		\includegraphics[width=12.12cm]{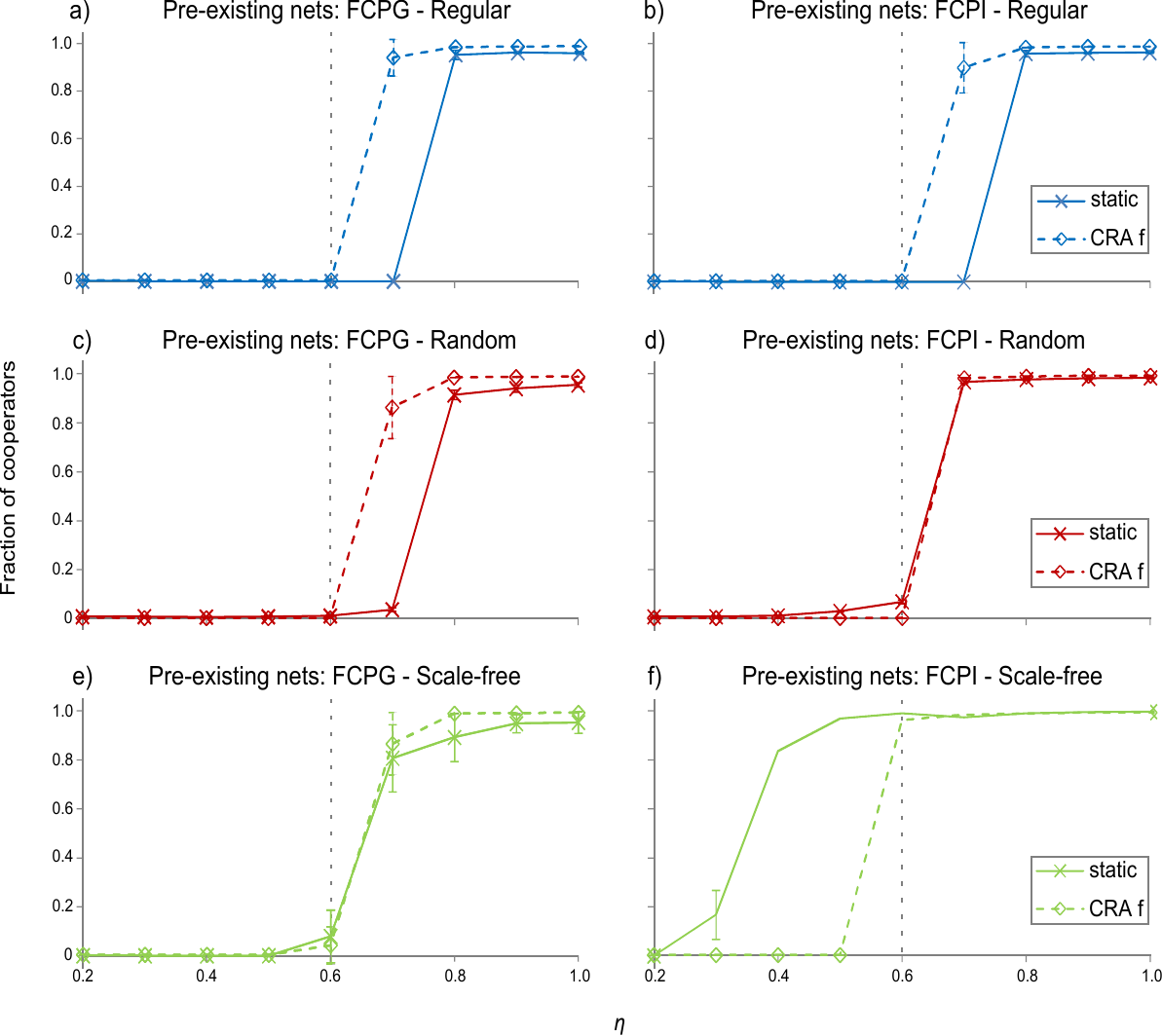}
		\caption{Plots comparing results from networks with static population size to those where size fluctuates around a nominal maximum value. Three types of pre-existing networks were investigated: regular, random and scale-free.   Final fraction of cooperators in population is plotted against $\eta$, the PGG reward multiplier $r$, normalised with respect to average neighbourhood size ($g = 5$). Variability (25 replicates) is indicated by error bars showing 95\% confidence intervals. Simulation details are as described in Methods section. The dashed line at $\eta = 0.6$ is a reference for the eye.}
		\label{fig:stat_v_fluc_for_PENs}
	\end{center}
\end{figure}

Figure \ref{fig:stat_v_fluc_for_PENs} presents our results so as to separately illustrate the impact of fluctuation on each of the network types studied. Here we generally see (comparing dashed line for all plots) that CRA-fluctuation shows increased or similar levels of cooperation in comparison to results for static networks. We observe this effect in all cases except FCPI initially scale-free networks (Fig. \ref{fig:stat_v_fluc_for_PENs}f).  

Our general observation here, that cooperation is increased or unchanged by fluctuation, is primarily a result of the CRA-fluctuation mechanism converting all networks to the same compressed exponential degree distribution, regardless of starting topology. Whilst the fluctuation model is in this way able to shift cooperation profiles to the left (increasing cooperation), such a change can only be achieved for networks having initially lower heterogeneity (random and regular networks).  Hence, in the case of scale-free (highly heterogeneous) networks, levels of cooperation should be lower in the fluctuating network than in the static network. Our results for scale-free networks are however not as clear cut as this simple explanation would suggest and we now provide further clarification of why this is so. 

In Fig. \ref{fig:stat_v_fluc_for_PENs}e, for FCPG PGG on initially scale-free networks, we see no reduction in levels of cooperation when the fluctuation model is applied. The reason we do not see such a reduction is because, as highlighted earlier, FCPG restricts the levels of cooperation that are achievable in heterogeneous networks.  This constraint limits levels of cooperation in the case of static networks and there is no reason to expect the fluctuation model to be able to overcome such a constraint.  We hence see similar results for both fluctuation and static models in FCPG PGG.  In Fig. \ref{fig:stat_v_fluc_for_PENs}f, we clearly see how the absence of such a constraint gives us the expected results.  Here, for the fluctating network, we see reduced levels of cooperation due to CRA-fluctuation reducing the heterogeneity of the network as it converts it from scale-free to compressed exponential degree distribution. 

In summary we make the general observation that fluctuation is more beneficial to cooperation in the case of FCPG. Whereas FCPI enables higher levels of cooperation to be achieved without the need for such `further assistance'.

\subsection{Simulations in Networks Grown from Founder Populations}

We now report on results of simulations grown from founder populations of either 3 cooperators, or 3 defectors. We compare two implementations of our model.  In the first (`non-fluctuating'), the population grows by a process of CRA until it reaches a maximum size, after which the network structure remains constant. In the second implementation (`fluctuating'), the network grows by means of CRA until it reaches a nominal maximum size, whereupon it is pruned and then allowed to regrow. This fluctuation cycle repeats thereafter, until the simulation ends. Strategy updating continues throughout the entirety of both implementations. As previously, we present results for both FCPG and FCPI variants of the PGG.

\subsubsection{Effect of PGG Variant in Networks Grown from Founder Populations.}

\begin{figure}[!h]
	\begin{center}
		\includegraphics[width=12.065cm]{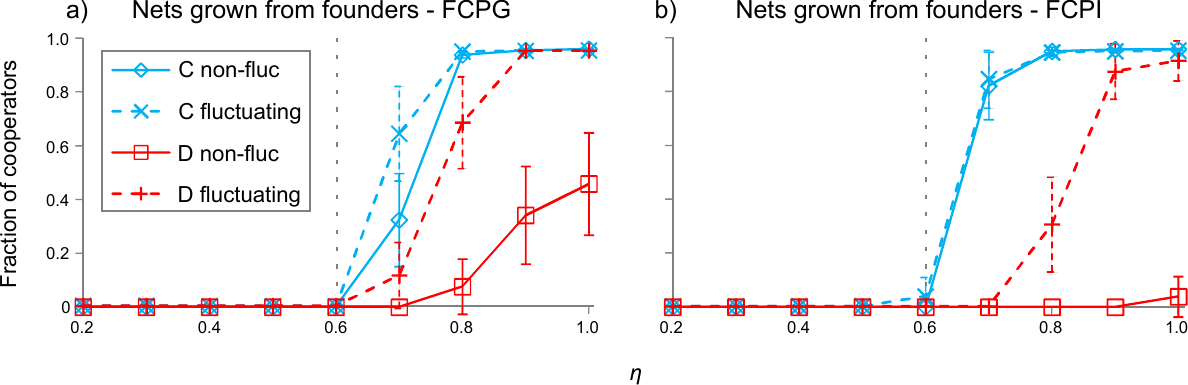}
		\caption{Plots comparing results from fluctuating and non-fluctuating simulations, for networks grown from founder populations (25 replicates). We compare networks grown from 3 cooperators to those grown from 3 defectors. Final fraction of cooperators in population is plotted against $\eta$, the PGG reward multiplier $r$, normalised with respect to average neighbourhood size ($g = 5$). Populations were specified to have a nominal maximum size of 1000 individuals, at which point fluctuation is triggered. Variability is indicated by error bars showing 95\% confidence intervals. Further details on simulations are as described in Methods section. The dashed line at $\eta = 0.6$ is a reference for the eye.}
		\label{fig:stat_v_fluc_for_seeds}
	\end{center}
\end{figure}

In the case of cooperator-founded populations, by comparing corresponding curves (cyan lines) between Figs. \ref{fig:stat_v_fluc_for_seeds}a and b, it appears that FCPI may result in a marginal increase in levels of cooperation compared to FCPG. Small increases would be consistent with our understanding that FCPI can relax the penalty paid by cooperator clusters in  heterogeneous networks. As described earlier, networks formed by CRA-fluctuation are only moderately heterogeneous, thus any increase afforded by FCPI over FCPG would be expected to be minor. 

In the case of defector-founded populations (red lines in Figs. \ref{fig:stat_v_fluc_for_seeds}a and b), we see what may be a minor difference between FCPG and FCPI in fluctuating networks (dashed lines) and a more marked difference for non-fluctuating networks (solid lines).  Curiously, FCPG \emph{in the case of defector founders} appears to \emph{promote} rather than restrict cooperation in non-fluctuating networks.  Closer inspection of the data for these specific FCPG simulations shows very high variability for values of $\eta > 0.8$.  Here individual replicate simulations proceeded to either high levels of cooperation or almost complete defection. We have been unable to establish a definitive explanation for this effect but strongly suspect that it is a result of FCPG favouring smaller neighbourhoods.  In defector-founded populations, simulations where many cooperators connect to the large `core' neighbourhood of defectors will only result in further growth of this defector core, since added cooperators are inevitably converted by the highly-connected defector core.  However in simulations where larger numbers of cooperators randomly attach elsewhere, it may become possible for initially small groups of cooperators to get a `toehold' in the population away from the main defector core. Growth of this type, where cooperation arises in smaller clusters away from the core, may be promoted by FCPG, which by penalising cooperators in larger neighbourhoods, favours those in smaller ones.

\subsubsection{Effect of Founder Population Strategy on Non-fluctuating and Fluctuating Networks.}

For cooperator-founded populations (blue lines in Figs. \ref{fig:stat_v_fluc_for_seeds}a and b), levels of cooperation are not markedly changed by incorporation of fluctuation once the network has reached its maximum size. We propose that the strategy assortativity that has taken place whilst the network was growing has maximised the amount of cooperation that can occur (for a given value of $\eta$).  Cooperation levels thus do not rise beyond this limit when fluctuation is added.  

For the more challenging, defector-founded scenario (red lines in Figs. \ref{fig:stat_v_fluc_for_seeds}a and b), fluctuation (dashed lines) brings about a marked increase in cooperation. To explain this effect we first consider the non-fluctuating system.  Here, an initial defector population converts all nodes that attach to it to defection and hence significantly biases the growing population against cooperation. The founder defectors develop increased connectivity over subsequent generations and become a well-connected core of defectors, easily capable of converting any new cooperators that may attach. In the non-fluctuating model, the starting point of defector founders thus `locks in' long term defector behaviour---to the extent that even where the dilemma collapses at $\eta = 1$, populations can still grow to be predominantly populated by defectors. In this scenario, rather like the growth of a coral reef, the core of defectors is `dead'.  `Life' (positive fitness values) only occurs at the periphery of the core where new cooperators  attach and hence allow defectors to gain positive fitness values.  If the new nodes are defectors they instantly become part of the core;  cooperators will do the same as soon as they have played PGG and been converted to defectors.  Thus the defector core continues to grow.    Whilst the core will primarily have zero fitness, it is still however impenetrable to invasion by cooperators, because they will be instantly converted as soon as they connect to the periphery of the core.

We now consider how the fluctuation model changes the above scenario to bring about such a marked increase in cooperation. Both fluctuating and non-fluctuating models operate identically until the maximum network size is reached. After this point, in the fluctuating model, the least fit individuals and the nodes they occupy are deleted from the population. New nodes are then added which link to randomly selected existing nodes and have randomly allocated strategies. Thus the defector core which is invulnerable in the non-fluctuating mode, contains many nodes which are highly vulnerable in the fluctuating model, due to their zero fitness. Our evolutionary model of fluctuating populations thus creates an escape from the domination of the defector core.   

We highlight that a preferential (rather than random) attachment system for network growth in a defector-founded population as we have described, may potentially \emph{reduce} the likelihood of cooperation emerging.  This is because in preferential attachment, new nodes would be most likely to connect with existing nodes of highest degree, i.e. the defector-founders. As a result any cooperators would be highly likely to be immediately converted to defectors.   Random attachment however allows for cooperators to connect elsewhere and hence allows for the development of cooperator clusters away from the founders where they are less likely to be converted to defection.

\section{Conclusion}

In this work we have developed a model of the coevolution of cooperative behaviour alongside the growth of a networked population. Importantly, our model demonstrates the emergence of such behaviour, in networks grown from non-cooperative founder members. Our results highlight that the absence of any perturbation of the system may potentially `lock in' defector behaviour in the long term---a concept that appears to merit further investigation. We also note the possibility that, in certain circumstances, preferential attachment can impede the emergence of network-reciprocal cooperation. In addition to investigating growing networks, we have  applied our model to pre-existing networks, populated with initially random strategies.  Regardless of initial topology, such networks achieve a compressed exponential degree distribution and the emergence of cooperation is observed. 

Our model has no requirements for agents to possess higher cognitive abilities such as memory or recognition. It also does not require underlying explanations to describe preferential attachment of nodes in network formation. Levels of network heterogeneity that are sufficient for cooperation to emerge, arise simply by random connections formed over time, combined with attrition of least fit members of the population.  Finally, we highlight that the model supports cooperation in cases of costly interaction (FCPG) and also where costs are trivialised (FCPI), real world scenarios being likely to lie somewhere along a spectrum between these two extrema. The fluctuation mechanism proposed here is the first example of a model describing the emergence of group-based cooperation in both growing \emph{and} dynamic-equilibrium networks.  As such it forms an important step in understanding the origins of cooperation in networks.

\subsubsection*{Acknowledgements}

This work has been funded by the Engineering and Physical Sciences Research Council (Grant reference number EP/I028099/1).

\bibliographystyle{splncs}
\bibliography{MILLER}

\end{document}